\documentclass[iop,apjl,tighten]{emulateapj}
\usepackage{amssymb}
\usepackage{times}
\usepackage{epsfig}
\usepackage{graphics}

\journalinfo{The Astrophysical Journal, 706: L?-L?, 2009}

\newcommand{\sax}{SAX J1808.4$-$3658}
\newcommand{\rin}{R_\mathrm{in}}

\slugcomment{Received 2009 September 3; accepted 2009 October 20; published 2009 ???}

\shorttitle{Pulse profile variations in accreting millisecond pulsars}
\shortauthors{Poutanen, Ibragimov \& Annala}

\begin{document}

\title{On the nature of pulse profile variations and timing noise \protect\\ in accreting millisecond pulsars}

\author{Juri Poutanen,\altaffilmark{1}  Askar Ibragimov,\altaffilmark{1,2,3} and Marja Annala\altaffilmark{1} }
\affil{$^1$ Department of Physics, Astronomy Division, PO Box 3000, FIN-90014 University of Oulu, Finland; 
juri.poutanen@oulu.fi \\ 
$^2$ Astronomy Department, Kazan State University, Kremlyovskaya 18, 420008 Kazan, Russia \\ 
$^3$ Faculty of Engineering and Natural Sciences, Sabanc{\i} University, Orhanl{\i}, Tuzla, 34956 \.{I}stanbul, Turkey}


\begin{abstract}
\noindent Timing noise in the data on accretion-powered millisecond pulsars (AMP) appears as irregular pulse phase jumps on timescales from hours to weeks. A  large systematic phase drift is also observed in the first discovered AMP  \sax.  To study the origin of these timing features, we use here the data of the well studied 2002 outburst of \sax.  We develop first a model for pulse profile formation accounting for the screening of the antipodal emitting spot by the accretion disk.  We demonstrate that the variations of the visibility of the antipodal spot associated with the receding accretion disk cause a systematic shift in Fourier phases, observed together with the changes in the pulse form. We show that a strong secondary maximum can be observed only in a narrow intervals of inner disk radii, which explains the very short appearance of the double-peaked profiles in  \sax.  By directly fitting  the pulse profile shapes with our model, we find that the main parameters of the emitting spot such as its mean latitude and longitude as well as the emissivity pattern change irregularly causing small shifts in pulse phase, and the strong profile variations are caused by the increasing inner disk radius. We finally notice that significant variations in the pulse profiles in the 2002 and 2008 outbursts of \sax\ happen at fluxes differing by a factor of 2, which can be explained if  the inner disk radius is not a simple function of the accretion rate, but depends on the previous history. 
\end{abstract}

\keywords{accretion, accretion disks -- methods: data analysis -- pulsars: individual (SAX J1808.4$-$3658) -- stars: neutron -- X-rays:  binaries} 

\section{Introduction}

The first accreting millisecond X-ray pulsar (hereafter AMP) \sax\ was discovered in 1998 \citep{WvdK98, CM98}. Since then, 12 objects of this class have been discovered, with the spin frequencies in the range 182--599 Hz. Many AMPs (e.g., IGR J00291+5934, XTE J1751$-$205, XTE J1807$-$294, \sax, XTE J1814$-$338) often show nearly sinusoidal profiles, consistent with only one visible  hotspot, probably because the accretion disk extends very close to the stellar surface. The general stability of the pulse profile allows to obtain a high photon statistics and to use the average pulse profile to get constraints on the neutron star mass-radius relation and the equation of state \citep{PG03}. 

However, sometimes the profiles do show variations. In \sax, for example, a dip appears around the pulse maximum at high photon fluxes, and at low fluxes the profile becomes skewed to the left and sometimes   is double-peaked. The pulse profile evolution  during its five observed outbursts is amazingly similar, with significant changes in the profiles accompanied by large (0.2 cycles) jumps  in the phase of the fundamental (see \citealt{BDM06,HPC08,HPC09,PW09}; \citealt{IP09}, hereafter IP09). 

Besides these noticeable profile changes, the timing noise with timescales from hours to weeks is known to be present in the data  of this and other AMPs, when the measured pulse phase delays show fluctuations around a mean trend  \citep{pap07,CCHY08,RDB08,PWvdK09}. There are many possible reasons for the pulse profile changes and the timing noise: changes in the spot shape and position, variations of the emission pattern, inner disk radius or the optical depth of the accretion stream \citep{p08}. Changes in the  position of the hotspots \citep{lamb08,PWvdK09} can cause random phase irregularities;  however, it is difficult to imagine that they can lead to qualitative  changes in the pulse form  (such as transformation of a single-peaked profile to a double-peaked)  unless accompanied by strong variation in the  emissivity pattern. 

On the other hand, the visibility of the secondary spot  might change even with small fluctuations of the accretion rate. IP09 proposed that the strong pulse profile variations in the 2002 outburst of \sax\ results from the appearance of the secondary antipodal spot to our view when the magnetospheric radius increases and the accretion disk recedes from the neutron star with the dropping accretion rate. The decreasing amplitude of Compton reflection in the spectra during the outburst supports this interpretation.

In this Letter, we directly fit the pulse profiles of \sax\ during its 2002 outburst with the theoretical model for AMP pulses including the effect of partial screening of the antipodal emitting spot by the accretion disk. We aim at  quantifying the variations in the position of the hotspot centroid, its emissivity pattern and the inner accretion disk radius, and determining the causes of the variations in the pulse shape and the  timing noise.

\begin{figure}
\centerline{\epsfig{file= 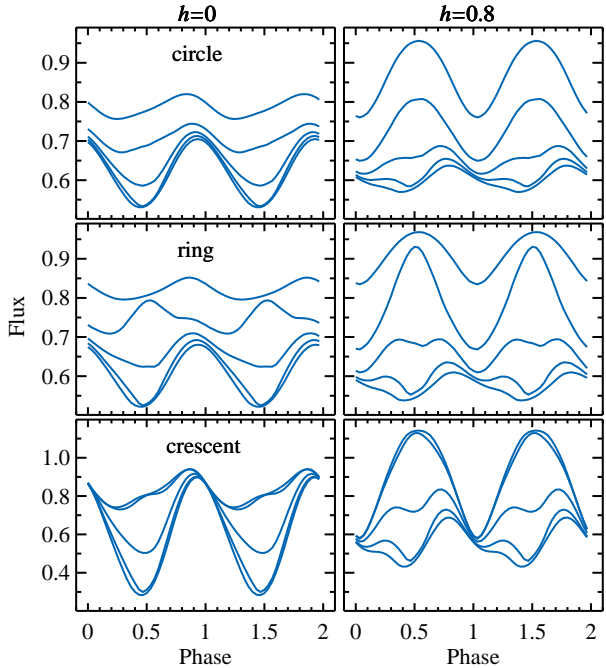,width=8cm}} 
\caption{Pulse profiles produced by two antipodal spots of various shapes for $\rin$ varying from 20 to 30 km with step 2.5 km (from bottom to top). 
The left hand side panels correspond to the blackbody emissivity pattern $h=0$, and the panels at the right are for $h=0.8$.  
At lowest radii, the view of antipodal spot is almost fully blocked by the disk, while at highest radii, the whole spot is visible.  
Stellar parameters: $M=1.4M_\odot$, $R=10.3$ km, $i=65\degr$, $\theta=10\degr$. The spot angular size $\rho$ is given by equation (\ref{eq:rhorin}), and for the ring and crescent geometries,  the inner radius is $\rho/\sqrt{2}$.  }
\label{f:profile}
\end{figure}

\section{Model}
\label{s:model}

We assume that the stellar magnetic dipole is displaced  by angle $\theta$ from  the rotational axis.  The accretion disk is disrupted by the stellar magnetic field at some radius $\rin$, which is a function of the accretion rate and the neutron star magnetic moment. Matter following magnetic field lines hits the stellar surface making two antipodal hot spots. For the dipole field, the outer spot edge is  displaced from the magnetic pole by an angle $\rho$: 
\begin{equation} \label{eq:rhorin}
\sin\rho= \cos \theta \sqrt{{R}/{\rin}}, 
\end{equation}
where $R$ is the neutron star radius. This gives us a rough estimation of the spot size as a function of $\rin$. 
The spot size and shape depend on the magnetic inclination and the inner disk radius in a complicated fashion. When 
$\theta\ll 1$ and $\rin\gg R$, the hotspot is nearly a circle. When the disk is close to the star, the spot becomes a ring-like because 
the matter accretes via field lines penetrating the disk  within a narrow ring. If the magnetic dipole is sufficiently inclined, the spot becomes a crescent-like \citep{RU04,KR05}. Thus we consider three simple geometries which incorporate all these cases: (1) a homogeneously emitting circle, (2) a ring, and (3) a crescent, which we approximate as a part of the ring.

For a given spot geometry and stellar parameters, we then compute the pulse profile. We account for general and special relativity effects (such as gravitational light bending, redshift,  aberration, Doppler boosting), and time delays. We also need to specify the  spectral shape and the angular emissivity pattern of radiation from the spot.  We follow the general methodology as described in detail in  \citet{PG03}, \citet{VP04}, and \citet{PB06}. As the  accretion disk might be rather close to the stellar surface, we  account also for the absorption by the disk as described in Appendix C of IP09.
We also approximate the theoretical profile by a sum of two harmonics: 
\begin{equation}\label{eq:cosines}
F(\phi)=\overline{F}\{ 1+a_1\cos[2\pi(\phi-\phi_1)] +a_2\cos[4\pi(\phi-\phi_2)] \} . 
\end{equation}

\begin{figure}
\centerline{\epsfig{file= 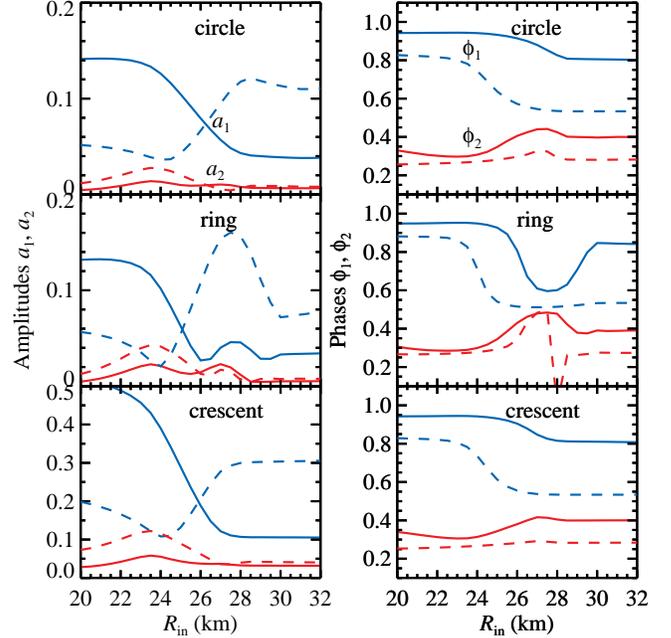,width=8.5cm}} 
\caption{Fourier amplitudes and phases of the fundamental and overtone (as given by Equation (\ref{eq:cosines})) of the pulse profiles shown in Figure \ref{f:profile} as   functions of the inner disk radius.  Solid and dashed  curves correspond to the cases of $h=0$ and $h=0.8$, respectively.  
}
\label{f:amplphase}
\end{figure}

For the illustration, we assume, rather arbitrarily, the following parameters: neutron star mass $M=1.4M_\odot$, radius $R=10.3$ km, spin frequency 401 Hz (corresponding to \sax), inclination $i=65\degr$, magnetic inclination $\theta=10\degr$, and a spot of angular radius $\rho$ given by Equation (\ref{eq:rhorin}).  
The spectral energy and angular distributions of the radiation intensity at the stellar surface are assumed to be represented by the expression 
\begin{equation}\label{eq:Ialpha}
I(E,\alpha)\propto\ (1-h\cos \alpha) \ E^{-(\Gamma-1)},  
\end{equation}
where $\Gamma\approx 2$ is the power-law photon spectral index observed in AMPs  and \sax\ in particular
\citep[see, e.g.][]{PG03,FBP05,FKP05,FPB07,p06}, $\alpha$ is the angle relative to the surface normal, and $h$ is the anisotropy parameter.  The assumed angular dependence with $h=0$ corresponds to the blackbody-like emission pattern, while $h\approx$ 0.6--0.8 is consistent with the thermal Comptonization in an optically thin slab, which is probably responsible for the power-law-like  spectra.

\begin{figure*}
\plotone{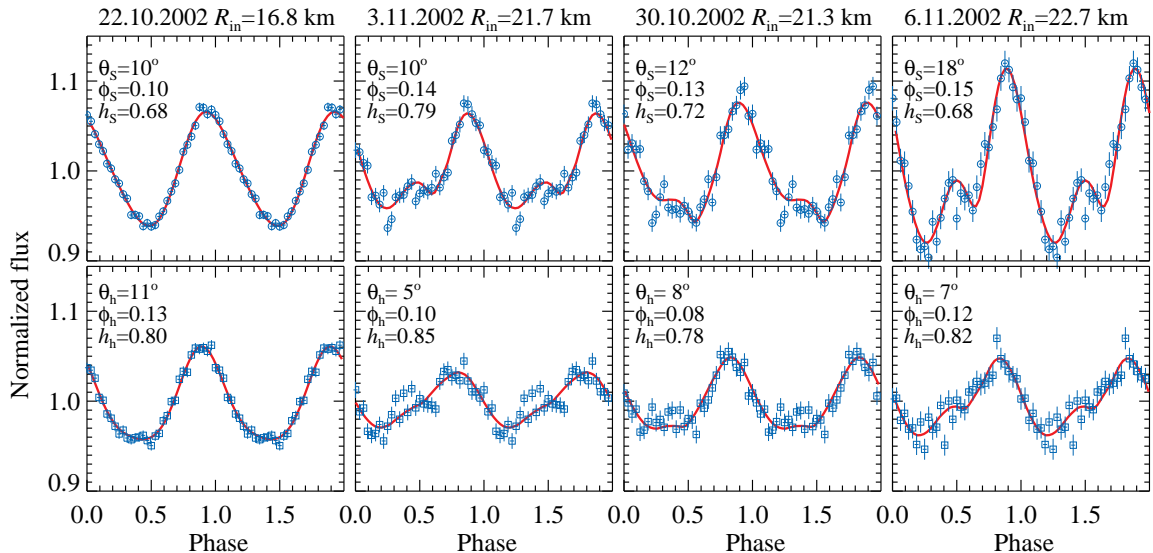}
\caption{Sample of pulse profiles of \sax\ observed during the October--November 2002 outburst, in order of decreasing flux. 
The  top and bottom panels correspond to the profiles in the 3.7--5.7  keV and  9.8--23.2 keV energy bands, respectively. 
The corresponding best-fit models parameters are indicated. The solid curves give the modeled pulse shapes. }
\label{f:pulses}
\end{figure*}

For a ring- and a crescent-like spots we assume the inner radius of the spot $\rho/\sqrt{2}$, and for the crescent we consider half a ring furthest away from the rotational pole similar to what is obtained in MHD simulations \citep{RU04,KR05}. Figure \ref{f:profile} shows the pulse profile variations   as a function of the inner disc radius for the three geometries and two anisotropy parameters $h=0,\ 0.8$. The noticeable signatures of the  second spot appear when  $\rin>20$ km,  the dramatic change in the pulse profile happens when the inner radius changes  from 24 to 26 km, as most of the antipodal spot then appears to our view. These fast changes are also reflected in a rather large  phase shift of the fundamental by $\sim$0.2 (see Figure \ref{f:amplphase}), but generally smaller variations of the phase of the overtone, similar to what is observed in \sax\ (see Figures 6 and 7 in IP09). The behavior of the amplitudes depends on the anisotropy.

While blackbody pulse shapes are rather smooth, the Comptonization emissivity pattern lead to the presence of narrower features and more pronounced secondary maxima; the ring generates sharper features in the lightcurves and more rapid evolution with $\rin$ comparing to the circle case. The crescent gives higher variability amplitude, because of its larger mean displacement with respect to the rotational pole of the star. However, the initial (when only one spot is visible) and the final (when both spots are visible) pulse profiles depend very little on the spot geometry, and changes of the Fourier phases with inner radius are also very similar. We also note that clearly double-peaked profiles are produced in a rather narrow range of $\rin$, when only a fraction of the secondary spot is visible. At large $\rin$, the profiles becomes again nearly sinusoidal in spite of the fact we observe both spots simultaneously.  

The phase shift of the fundamental shown in Figure \ref{f:amplphase} is a very robust prediction independent of the stellar parameters,
emissivity pattern, and spot shape. The exact disk radius when this occurs depends on the stellar compactness and observer inclination (see IP09). For a sufficiently small  $i$ and a not very compact star, the antipodal spot might never be visible. In that case, there will be no noticeable changes in the pulse profile with the accretion rate.

\begin{deluxetable*}{llcrrcccccc}
\tabletypesize{\scriptsize}
\tablecaption{Best Fit Parameters \label{tbl-1}}
\tablewidth{0pt}
\tablehead{
\colhead{MJD} & \colhead{Date} & \colhead{$F$\tablenotemark{a} } & \colhead{$\theta_\mathrm{s}$} & \colhead{$\theta_\mathrm{h}$  } &  \colhead{$\phi_\mathrm{s}$ } & \colhead{$\phi_\mathrm{h}$ } & \colhead{$h_\mathrm{s}$ } & \colhead{$h_\mathrm{h}$ }
& \colhead{ $\rin$ } & \colhead{$\chi^2/\mathrm{d.o.f.}$ }   \\
 & &  &  (deg) &  (deg) &   &  &  &   & (km)  &  } 
\startdata
52569.0--52569.5 & October 22  & 15.6 & 10.2 & 10.8 &  0.10 & 0.13 &  0.68 &  0.80 &  16.8 &  34/56 \\
52580.8--52581.5 & November 3 & 4.2 & 9.7 & 5.2   &  0.14 & 0.10 &  0.79 &  0.85 & 21.7 &  80/56  \\
52581.6--52582.0 & November 3  & 3.6 & 8.2 & 5.4   & 0.15 & 0.06 &   0.75 &  0.83 &  22.5 &  62/56  \\
52576.9--52577.4 & October 30 & 3.5  & 11.6 & 8.1  & 0.13 & 0.08 &   0.72 &  0.78 &  21.3 &  86/56  \\
52577.9--52578.1 & October 31 & 2.4  & 18.4 & 6.6  & 0.13 & 0.09 &   0.67 &  0.84 & 22.8 &  92/56  \\
52583.8--52584.2 & November 6 & 1.9 & 18.1 & 7.0  & 0.15 & 0.12 &   0.68 &  0.82 & 22.7 &  61/56  \\
52578.9--52579.5 & November 1 & 1.5 & 19.3 & 5.7  &  0.12 & 0.11 &  0.61 &  0.79 &  23.2 &  74/56   
\enddata
\tablenotetext{a}{Flux in 2.5--25 keV band in units $ 10^{-10} $ erg $\mathrm{cm}^{-2}$ s$^{-1}$.}
\end{deluxetable*}

\section{Comparison with the observed profile evolution}
\label{s:obsc}

Now we make a detailed comparison of the theoretical pulse profiles predicted by the model to the data on 2002 outburst of  \sax\ well covered by the {\it Rossi X-ray Timing Explorer}.  The data and their reduction  are described in details in IP09.  At the peak stage  (see \citealt{HPC08} and IP09 for the definition of outburst stages), profiles are nearly sinusoidal with a dip  noticeable at the pulse maximum at energies above 10 keV. This might be caused by  absorption in the accretion column (see IP09) complicating the pulse modeling, and therefore we do not consider this stage in the paper. 
During the slow decay stage on October 17--27,  the pulse profile, slightly skewed to the right (see Figure \ref{f:pulses}, left panels), was extremely stable, in spite of the fact that the flux was changing by a factor of 4. At the rapid drop stage (October 27 -- November 1) the flux fell down rapidly, while the pulse profile was evolving: on October 29 the secondary maximum -- a signature of the second hot spot -- becomes pronounced and after October 30 the pulse became skewed to the left and the secondary maximum shifted to the rising part of the main peak. In Figure \ref{f:pulses}, we present a subset of the observed pulse shapes. 

The pulse profiles can be rather well described by  a sum of two Fourier harmonics, fundamental and overtone, given by Equation (\ref{eq:cosines}).  The dramatic variability of the pulse profiles is reflected in a phase shift of the fundamental by 0.2 observed within a couple of days around October 30, while the phase of the overtone did not change significantly  (\citealt{BDM06,HPC08}; IP09).  The behavior of the pulse phases and amplitudes varies with energy: for example,  the amplitude of the high-energy pulse decreases with time, while the pulsation amplitude at low energies noticeably increases (IP09). The reason for that most probably lies in the fact that the soft blackbody which contributes significantly to the total spectrum below $\sim$7 keV  (\citealt{GDB02,PG03}; IP09) might be produced in a region displaced from the accretion shock where harder X-rays are produced.

For comparison of the model to the data, we take the same stellar parameters as given in Section \ref{s:model}, inclination $i=60\degr$,  and consider a filled circle geometry of the spot. We fit simultaneously the pulse profiles in two broad energy bands 3.7--5.7 and  9.8--23.2 keV. The second band is dominated by the Comptonized components, while the blackbody contributes at the lower-energy band. We consider a possibility that photons at these energies can be produced predominantly at different positions, determined by the colatitude $\theta_\mathrm{s,h}$ and the azimuth $\phi_\mathrm{s,h}$ (in units of $2\pi$) of the spot center.  The angular emissivities of the radiation in the two bands are described by Equation (\ref{eq:Ialpha}) with parameters $h_\mathrm{s,h}$. The inner disk radius $\rin$ is of course the same for both energies at each moment of the outburst, and the spot size is given by Equation  (\ref{eq:rhorin}) with $\theta=0$. The best-fit parameters are given in Table \ref{tbl-1} and the representative pulse profiles are shown in Figure \ref{f:pulses}.
 
We see that the model describes the pulse profiles rather well, except some narrow features. In general, the resulting $\rin$ increases when the flux decreases, but not monotonically. We also see some variations in positions of the spot centroids, both in latitude and in longitude. These changes can be associated with the timing noise (i.e., erratic variations of the Fourier phases). The jump in the fundamental phase observed around October 30, which happened together with the dramatic change in the pulse profile, is explained here by the appearance of the antipodal spot when the inner disk radius is about 20 km (the exact value would depend on the stellar compactness and the inclination). The origin of some discrepancies between the model and the data probably lies in a simplified spot shape assumed here (the pulse profiles depend  considerably on that, see  Section \ref{s:model}) and in the assumption of the equal area spots producing emission at different energies. 

Since the evolution of the pulse profiles during various outbursts of \sax\ is very similar, our model thus can be applied directly to other outbursts too. It is worth noting, however, that the changes in the pulse profile and appearance of double-peaked profile in the 2008 data happen at a flux twice as large as in the 2002 and 2005 outbursts \citep[see][]{HPC08,HPC09}. This might imply that  the inner disk radius and the magnetospheric radius are not a simple function of the accretion rate, but also of the history of the outburst. We can speculate that in 2008, a smaller total accreted mass resulted in a less effective reduction of the stellar magnetic field  \citep{cum08} and, therefore, the same inner disk radius could have been reached at a higher accretion rate.  
The phase fluctuations correlated with the flux as observed, for example, in XTE J1807$-$294 and XTE J1814$-$338 \citep{pap07,CCHY08,RDB08,PWvdK09}, cannot be easily explained by the inner disk variations. 

\section{Conclusions}

We have developed a model for the pulse profile formation in AMPs which accounts for the partial screening of the antipodal emitting spot by the accretion disk. We have demonstrated that the appearance of the antipodal spot, due to the increasing inner disk radius, leads to sharp changes in the pulse profile and corresponding jump in Fourier phases. We showed that a strong secondary maximum appears only in a rather narrow intervals of inner disk radii, explaining the very short appearance of the double-peaked profiles in  \sax. 

By directly fitting the model to the pulse shapes of \sax\ observed during its 2002 outburst, we were able to quantify the variations in the position of the spot centroids, their emissivity pattern and the inner accretion disk radius. The sharp jumps in Fourier phases, observed together with the strong changes in the pulse form,  can be explained by the variations of the visibility of the antipodal spot associated with the receding accretion disk.  We also note that a factor of 2 difference in fluxes in 2002 and 2008 outbursts of \sax, when the pulse profile started to change significantly, implies that the inner disk radius is not a simple function of the accretion rate, but might depend on the previous history. 
We finally note that  many physical  timing noise mechanisms probably operate in AMPs simultaneously.

\acknowledgments

J.P. acknowledges support from the Academy of Finland grants 110792 and 127512.  A.I. was supported by the Finnish Graduate School in Astronomy and Space Physics, the  V\"ais\"al\"a and Kordelini Foundations, and the Russian Presidential program for support of leading science schools (grant NSh 4224.2008.2).  M.A. is thankful to the  V\"ais\"al\"a Foundation.
We also acknowledge the support of the International Space Science Institute (Bern, Switzerland).


\begin{thebibliography}{20}
\expandafter\ifx\csname natexlab\endcsname\relax\def\natexlab#1{#1}\fi

\bibitem[{{Burderi} {et~al.}(2006){Burderi}, {Di Salvo}, {Menna}, {Riggio}, \&  {Papitto}}]{BDM06}
{Burderi}, L., {Di Salvo}, T., {Menna}, M.~T., {Riggio}, A., \& {Papitto}, A.
  2006, \apjl, 653, L133

\bibitem[{{Chakrabarty} \& {Morgan}(1998)}]{CM98}
{Chakrabarty}, D. \& {Morgan}, E.~H. 1998, \nat, 394, 346

\bibitem[{{Chou} {et~al.}(2008){Chou}, {Chung}, {Hu}, {Yang}, \& {Papitto}}]{CCHY08}
  {Chou}, Y., {Chung}, Y., {Hu}, C.-P., \& {Yang}, T.-C. 
  2008, \apj, 678, 1316
  
\bibitem[{{Gierli{\'n}ski} {et~al.}(2002){Gierli{\'n}ski}, {Done}, \&  {Barret}}]{GDB02}
{Gierli{\'n}ski}, M., {Done}, C., \& {Barret}, D. 2002, \mnras, 331, 141

\bibitem[{{Cumming}(2008)}]{cum08}
{Cumming}, A. 2008, in AIP Conf. Ser. 1068, 
A Decade of Accreting X-ray Millisecond Pulsars, 
 ed. R.~{Wijnands} et al. (Melville, NY: AIP), 152

\bibitem[{Falanga} {et~al.}(2005{\natexlab{a}})]{FBP05}
{Falanga}, M. et al., 2005{\natexlab{a}}, \aap, 436, 647

\bibitem[{Falanga} {et~al.}(2005{\natexlab{b}})]{FKP05}
{Falanga}, M. et al., 2005{\natexlab{b}}, \aap, 444, 15

\bibitem[{Falanga} {et~al.}(2007)]{FPB07}
{Falanga}, M., {Poutanen}, J., {Bonning}, E.~W., {Kuiper}, L., 
	{Bonnet-Bidaud}, J.~M., {Goldwurm}, A., {Hermsen}, W. \&  	{Stella}, L.,
 2007, \aap, 464, 1069

\bibitem[{{Hartman} {et~al.}(2008){Hartman}, {Patruno}, {Chakrabarty},
  {Kaplan}, {Markwardt}, {Morgan}, {Ray}, {van der Klis}, \&  {Wijnands}}]{HPC08}
{Hartman}, J.~M. et al., 
  2008, \apj, 675, 1468

\bibitem[{{Hartman} {et~al.}(2009){Hartman}, {Patruno}, {Chakrabarty},
  {Markwardt}, {Morgan}, {van der Klis}, \& {Wijnands}}]{HPC09}
{Hartman}, J.~M., {Patruno}, A., {Chakrabarty}, D., {Markwardt}, C.~B.,
  {Morgan}, E.~H., {van der Klis}, M., \& {Wijnands}, R. 2009, ApJ, 702, 1673

\bibitem[{{Ibragimov} \& {Poutanen}(2009)}]{IP09}
{Ibragimov}, A. \& {Poutanen}, J. 2009, \mnras, in press  (arXiv:0901.0073)  (IP09) 

\bibitem[{{Kulkarni} \& {Romanova}(2005)}]{KR05}
{Kulkarni}, A.~K. \& {Romanova}, M.~M. 2005, \apj, 633, 349

\bibitem[{{Lamb} {et~al.}(2008){Lamb}, {Boutloukos}, {Van Wassenhove},
  {Chamberlain}, {Lo}, {Clare}, {Yu}, \& {Miller}}]{lamb08}
{Lamb}, F.~K., {Boutloukos}, S., {Van Wassenhove}, S., {Chamberlain}, R.~T.,
  {Lo}, K.~H., {Clare}, A., {Yu}, W., \& {Miller}, M.~C. 2008,  arXiv:0808.4159
  
\bibitem[{{Papitto} {et~al.}(2007){Papitto}, {di Salvo}, {Burderi}, {Menna},  {Lavagetto}, \& {Riggio}}]{pap07}
{Papitto}, A., {di Salvo}, T., {Burderi}, L., {Menna}, M.~T., {Lavagetto}, G.,
  \& {Riggio}, A. 2007, \mnras, 375, 971

\bibitem[{{Patruno} {et~al.}(2009{\natexlab{a}}){Patruno}, {Watts}, {Klein-Wolt}, {Wijnands}, \& {van der Klis}}]{PW09}
{Patruno}, A., {Watts}, A.~L., {Klein-Wolt}, M., {Wijnands}, R., \& {van der
  Klis}, M. 2009{\natexlab{a}},  arXiv:0904.0560

\bibitem[{{Patruno} {et~al.}(2009{\natexlab{b}}){Patruno}, {Wijnands}, \& {van
  der Klis}}]{PWvdK09}
{Patruno}, A., {Wijnands}, R., \& {van der Klis}, M. 2009{\natexlab{b}}, \apjl,
  698, L60
  
\bibitem[{{Poutanen}(2006)}]{p06}
{Poutanen}, J. 2006, Adv. Space Res., 38, 2697

\bibitem[{{Poutanen}(2008)}]{p08}
{Poutanen}, J. 2008, in AIP Conf. Ser. 1068, 
A Decade of Accreting X-ray Millisecond Pulsars, 
 ed. R.~{Wijnands} et al. (Melville, NY: AIP),  77

\bibitem[{{Poutanen} \& {Beloborodov}(2006)}]{PB06}
{Poutanen}, J. \& {Beloborodov}, A.~M. 2006, \mnras, 373, 836

\bibitem[{{Poutanen} \& {Gierli{\'n}ski}(2003)}]{PG03}
{Poutanen}, J. \& {Gierli{\'n}ski}, M. 2003, \mnras, 343, 1301

\bibitem[{{Riggio} {et~al.}(2008){Riggio}, {Di Salvo}, {Burderi}, {Menna},  {Papitto}, {Iaria}, \& {Lavagetto}}]{RDB08}
{Riggio}, A., {Di Salvo}, T., {Burderi}, L., {Menna}, M.~T., {Papitto}, A.,
  {Iaria}, R., \& {Lavagetto}, G. 2008, \apj, 678, 1273

\bibitem[{{Romanova} {et~al.}(2004){Romanova}, {Ustyugova}, {Koldoba}, \&  {Lovelace}}]{RU04}
{Romanova}, M.~M., {Ustyugova}, G.~V., {Koldoba}, A.~V., \& {Lovelace},
  R.~V.~E. 2004, \apj, 610, 920

\bibitem[{{Viironen} \& {Poutanen}(2004)}]{VP04}
{Viironen}, K. \& {Poutanen}, J. 2004, \aap, 426, 985

\bibitem[{{Wijnands} \& {van der Klis}(1998)}]{WvdK98}
{Wijnands}, R. \& {van der Klis}, M. 1998, \nat, 394, 344

\end{thebibliography}
\end{document}